\documentclass[aip,amsmath,amssymb,graphicx,superscriptaddress,thelongbibliography]{revtex4-1}
\draft
\usepackage{graphicx}
\usepackage{dcolumn}
\usepackage{bm}
\usepackage{color}
\usepackage{float}
\usepackage{mathptmx}
\begin{document}

\title[Chemotaxis of chiral swimmers]{External chemical gradient leads to efficient swimming of chiral swimmers}

\author{Ruma Maity}
\affiliation{Department of Physics, Indian Institute Science Education and Research Bhopal, Bhopal 462066, India}

\author{P.S. Burada}\email{psburada@phy.iitkgp.ac.in}
\affiliation{Department of Physics, Indian Institute of Technology Kharagpur, Kharagpur 721302, India}

\begin{abstract}

External gradients can strongly influence the collective behavior of microswimmers. In this paper, we study the behavior of two hydrodynamically interacting self-propelled chiral swimmers, in the low-Reynolds number regime, under the influence of an external linear chemical gradient. We use the generalized squirmer model called the chiral squirmer, a spherically shaped body with an asymmetric surface slip velocity, to represent the swimmer. We find that the external gradient favors the attraction between the swimmers and, in some situations, leads to a bounded state in which the swimmers move in a highly synchronous manner. Further, due to this cooperative motion, swimmers efficiently reach the chemical target compared to the individual swimmers. This study may help to understand the collective behavior of chiral swimmers and to design synthetic microswimmers for targeted drug delivery. 
\end{abstract} 
\maketitle

\section{Introduction}
\label{Intro}

Microorganisms are tiny, but their collective behavior can influence the climate and human life in various ways. For example, massive plankton blooms in the ocean, harmful red tides along the coastline, and bioconvection \cite{pallat}. Individual microorganisms exhibit various propulsive mechanisms depending on their size, shape, etc. \cite{Eric_book}. When microorganisms, e.g., {\it E-Coli}, sperm cells, {\it Paramesium}, and {\it Volvox}, swim in a fluid, the viscous force from the surrounding medium dominates over the inertia of the body \cite{purcell}. The corresponding Reynolds number is nearly zero for these microswimmers. The hydrodynamic interaction among the microswimmers may lead to long-range rotational order \cite{simha1, ramaswamy1}. It is due to the existence of dynamical instability in the system. Note that it is possible to stabilize the instability by various means, for example, chemical signaling \cite{nejad}. Biofilm formation is an apt example of collective sensing of chemical stimulus by a bacteria colony \cite{stoodley}.

A popular squirmer model \cite{lighthill, blake}, a sphere with an axisymmetric surface slip, is used to understand the hydrodynamic behavior of spherical or nearly spherical-shaped microorganisms. Note that the squirmer can exhibit translational motion only. However, in general, many microorganisms exhibit translational and rotational motion. Resulting in swimming in helical paths and exhibiting chiral flows \cite{crenshaw, burada}. Notably, the non-axisymmetric flow feature of a swimmer is common in literature \cite{pak, ghose, felderhof}. Therefore, the generalized squirmer model called the chiral squirmer \cite{burada,pak,burada_lub} is more applicable for studying the collective behavior of spherical self-propelled bodies. In the chiral squirmer model, the tangential slip velocity on the body's surface considers both the body's translational and rotational degree of freedom. Also, note that a pair of chiral swimmers exhibit a peculiar bounded state, in addition to commonly observed states such as convergence and divergence states, due to the non-axisymmetric nature of the body as reported in our earlier works \cite{burada, burada_lub}. When the separation distance between the swimmers is significant compared to the swimmer's size, the swimmer's flow field affects the other swimmers' migration velocity and vice versa. However, when the swimmers approach each other, the lubrication forces and torques arise and control the hydrodynamic behavior of the swimmers \cite{wang,di,burada_lub}. 

In general, microorganisms sense the external stimulus in their vicinity and respond to it by moving towards \cite{kirchman} or away from it. 
This directional movement in the presence of external chemical stimulus is known as chemotaxis. It plays a vital role in biological pattern formation \cite{saintillan}, altering the viscosity of the medium \cite{sokolov, haines}, generating turbulent flows \cite{dunkel}, etc. Chemotaxis of sperm cells, {\it E. coli}, {\it Dictyostelium}, {\it Paramecium}, {\it Tetrahymena thermophila}, {\it Amoeba proteus}, etc. is ubiquitous in nature \cite{Larsen, Jikeli, Sarvestani}. Artificial swimmers are also able to exhibit chemotaxis under controlled experimental conditions. Unlike the swimming strategies of the living microorganisms, different mechanisms like the Marangoni effect \cite{jin}, the active diffusion \cite{hong}, the interfacial tension \cite{paxton}, and the extraction of work from active medium \cite{geiseler} prevail for artificial swimmers to respond to the chemical gradient. 

In this article, we study the response of a pair of chiral swimmers to the external chemical gradient by considering the near- and far-field hydrodynamic interaction between them. The strength of the gradient solely controls the response of a single swimmer to the external gradient. However, for a pair of swimmers, the hydrodynamic interaction between them plays a crucial role in the chemotaxis of the swimmers. The latter plays a constructive role in the success of chemotaxis. The paper is organized as follows. The general chiral squirmer model is briefly discussed in section~\ref{sec:model}. In section \ref{sec:no-noise}, we analyze the hydrodynamic behavior of two chiral swimmers in the presence of a linear chemical gradient. The main conclusions are provided in section~\ref{sec:conclusions}.

\section{Hydrodynamics of a chiral squirmer}
\label{sec:model}

The low Reynolds number swimmers \cite{lighthill,purcell} obey the Stokes equation \cite{happel} given by
\begin{equation}
 \eta\nabla^2 \mathbf{u} = \nabla p\,,
 \label{eq:stokes}
\end{equation}
where $\eta$ is the viscosity of the fluid around the body, 
$\mathbf{u}$ is the velocity field, and $p$ is the corresponding pressure field 
which plays the role of a Lagrange multiplier to impose the incompressibility 
constraint $\nabla \cdot \mathbf{u} = 0$. In the chiral squirmer model, 
the effective slip velocity \cite{burada} on the surface of a spherical 
non-deformable body of radius $a$ is prescribed by
\begin{align}
\mathbf{u}_s = \displaystyle{\sum\limits_{l=1}^\infty}\,\displaystyle{\sum\limits_{m= -l}^l}
\Big[-\beta_{lm}\, {\boldsymbol \nabla}_s 
\left( P_l^m(\cos\theta) \,e^{i m \phi} \right)   
+ \gamma_{lm}\, {\bf \hat{r}} \times {\boldsymbol \nabla}_s \left( P_l^m(\cos\theta) \,e^{i m \phi} \right) \Big]\,,
\label{eq:slip}
\end{align}
where $\nabla_s$ = $\mathbf{e_\theta} {\partial}/{\partial \theta} + \mathbf{e_ \phi} ({1}/{\sin \theta})({\partial}/{\partial \phi})$ is the surface gradient operator, 
${\bf \hat{r}}$ is the unit vector in radial direction, $P_l^m(\cos\theta) \,e^{i m \phi}$
are non-normalized spherical harmonics, where $P_l^m(\cos\theta)$ 
are the associated Legendre polynomials of order $m$ and degree $l$. 
The complex coefficients $\beta_{l m}$ and $\gamma_{l m}$ are the surface slip velocity mode amplitudes. 
We introduce the real and imaginary parts of these amplitudes 
as $\beta_{l m} = \beta_{l m}^r + i\,m\, \beta_{l m}^i$ and $\gamma_{l m} = \gamma_{l m}^r + i\,m\, \gamma_{l m}^i$ with complex conjugates $\beta_{l m}^\ast = (-1)^m \beta_{l, -m}$ and $\gamma_{l m}^\ast = (-1)^m \gamma_{l, -m}$, respectively. 

\begin{figure}
\includegraphics[scale=0.3]{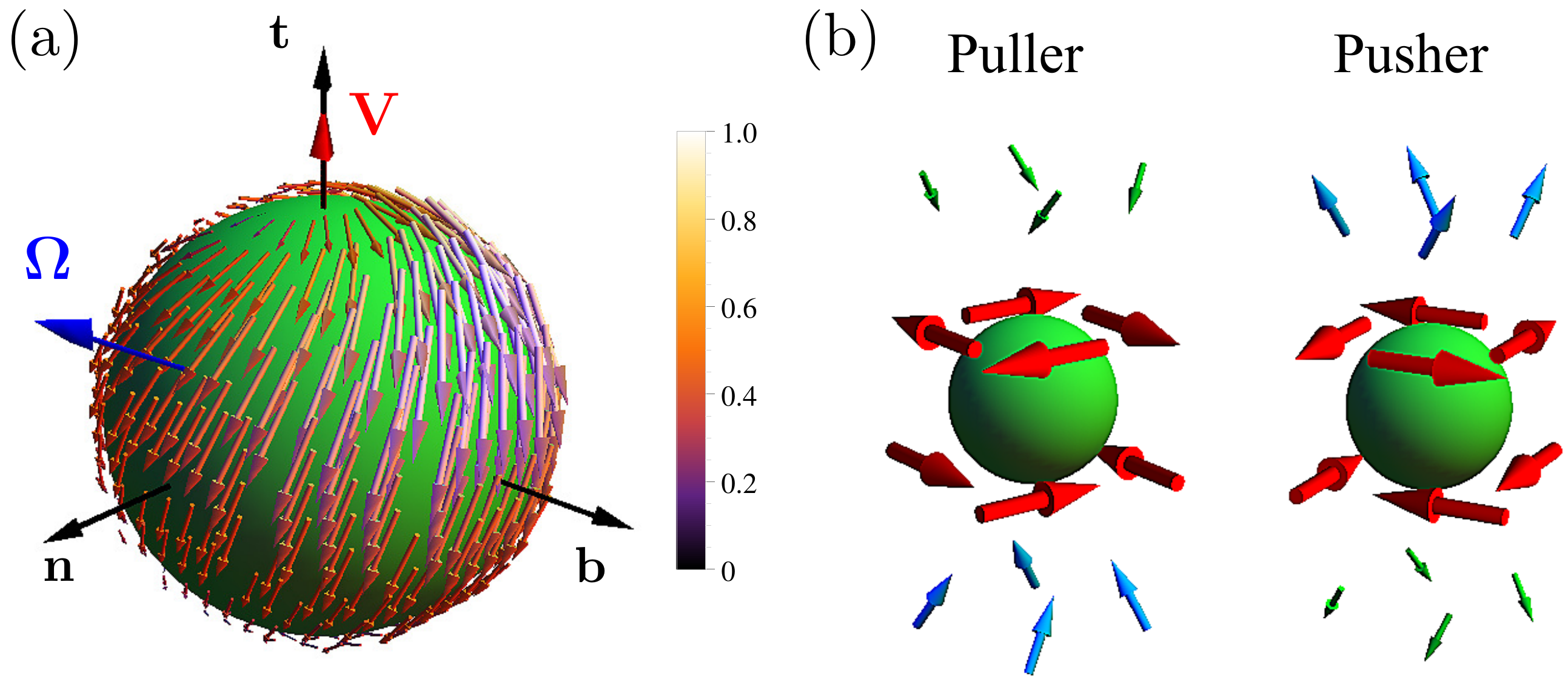}
\caption{\label{fig:CS}
Chiral squirmer (a) with a prescribed surface slip velocity in the body-fixed reference frame ${\bf n}$, ${\bf b}$, and ${\bf t}$. The corresponding velocity and rotation of the swimmer are $\mathbf{V}$ and $\bm{\Omega}$, respectively. The types of swimmers are shown in (b). For the puller-type swimmer $\beta_{20}^r > 0$ and for the pusher-type swimmer $\beta_{20}^r< 0$ (see the text for details). $\beta_{20}^r = 0$ corresponds to the neutral swimmer (not shown).}
\end{figure}

Interestingly, the swimming velocity $\mathbf{V}$ and the rotation rate $\bm{\Omega}$ of the swimmer can be obtained directly from the surface slip velocity Eq.~\ref{eq:slip}, using the reciprocal theorem, without explicitly calculating the corresponding flow- and stress-field (see Ref.\cite{stone}).
They can be obtained in the body-fixed reference frame as 
$\mathbf{V} = 2(\beta_{11}^r , \, \beta_{11}^i , \, \beta_{10}^r )/3$ and
${\boldsymbol \Omega} = (\gamma_{11}^{\,r}, \, \gamma_{11}^{\,i}, \, \gamma_{10}^{\,r} )/a$ \cite{burada}. 
Without loss of generality, the frame ${\bf n}, {\bf b}$, and ${\bf t}$ can be chosen such that ${\bf t}$ points in the direction of velocity. Accordingly, we have $\beta_{11}^r = \beta_{11}^i = 0$ and write $\beta_{1 0}^r = 3 v/2$ such that $v = |\mathbf{V}|$ is the speed of the swimmer.
Thus, the velocity and rotation rate of the chiral swimmer read \cite{burada},
\begin{align}
\mathbf{V} & = v \,  {\bf t} \ , \\ 
\label{eq:vel_rot}
{\boldsymbol \Omega} & =  \frac{\gamma_{11}^{\,r}}{a} \,{\bf n} +  \frac{\gamma_{11}^{\,i}}{a} \,{\bf b} +  \frac{\gamma_{10}^{\,r}}{a} \, {\bf t} \,.
\end{align} 
Also, for simplicity, we choose that the swimmer has the rotation rate in the ${\bf n-t}$ plane only. 
With this choice, we have $\gamma_{11}^{\,i} = 0$.
In addition, we choose the magnitude of the rotation as 
$|{\boldsymbol \Omega}| = v/a$ such that the components of the rotation rate can be expressed as 
$\gamma_{11}^{\,r}/a = (v/a)\sin \chi \,\, , \gamma_{11}^{\,i}/a = 0$ and $\gamma_{10}^{\,r}/a = (v/a)\cos \chi$, where $\chi$ is the angle between $\mathbf{V}$ and ${\boldsymbol \Omega}$. 
Fig.~\ref{fig:CS}(a) depicts the chiral squirmer with non-axisymmetric surface slip (Eq.~\ref{eq:slip}), its swimming direction, and the rotation rate. 
The corresponding equations of motion of the swimmer can be obtained 
using the force- and torque-balance conditions as 
\begin{align}
\dot{\textbf{q}} = \mathbf{V}, \,\,\,\, 
\dot{\textbf{n}} = \bm{\Omega}\times \textbf{n},\,\,\,\,
\dot{\textbf{b}} = \bm{\Omega}\times \textbf{b},\,\,\,\,
\dot{\textbf{t}} = \bm{\Omega}\times \textbf{t}\,,
\label{eqn:single}
\end{align}
where $\textbf{q}$ is the swimmer's position, and the dot represents the derivative with respect to time.
For $\mathbf{V} \parallel \bm{\Omega}$, we get $\chi = 0$, and the resulting swimming path is a straight line. 
In this case, the swimmer rotates around the axis of motion. 
For $\chi = \pi/2$, the chiral swimmer moves in a circular path in a plane. 
For other values of $\chi$, the path of the chiral swimmer is helix (see Refs. \cite{burada, burada_lub}).

Using the boundary conditions, $\mathbf{u} = \mathbf{u}_s$ at the surface of the chiral squirmer and $\mathbf{u} \to 0$ as $\mathbf{r} \to \infty$, i.e., in the laboratory frame of reference, the Stokes equation (Eq.~\ref{eq:stokes}) can be solved analytically to obtain the corresponding flow- and pressure-field \cite{burada,burada_lub}. They read,  
\begin{align}
\label{eq:VelocityField}
{\bf u}_\mathrm{lf}({\bf r})  
   & = \frac{3 v}{2} \frac{a^3}{r^3} 
   \left[ P_1({\bf t} \cdot {\bf \hat{r}})\,{\bf \hat{r}} - \frac{\bf t}{3} 
   \right]   
+ 3\,\beta_{2 0}^r \left( \frac{a^4}{r^4} - \frac{a^2}{r^2} \right)\,
P_2({\bf t} \cdot {\bf \hat{r}})\, {\bf \hat{r}}  \nonumber \\
& + \beta_{2 0}^r \frac{a^4}{r^4} 
P_2^{\prime}\left( {\bf t} \cdot {\bf \hat{r}} \right) [ ({\bf t} \cdot {\bf \hat{r}}) {\bf \hat{r}} - {\bf t} ] 
- \gamma_{2 0}^{\,r}\,\frac{a^3}{r^3} \, 
P_2^{\prime}\left( {\bf t} \cdot {\bf \hat{r}} \right)
{\bf t} \times {\bf \hat{r}} \,,\\ 
\label{eq:pressure_LB}
  p_\mathrm{lf}({\bf r})
  & =  - 2 \eta\,\beta_{2 0}^r \,\frac{a^2}{r^3} \,P_2\left({\bf t} \cdot {\bf \hat{r}}\right) \,,
\end{align}
where 
${\bf t}$ is the swimming direction, 
$r$ is the distance from the center of the swimmer where the flow field is determined, 
${\bf \hat{r}} = {\bf r}/r$ is the unit radial vector, 
$P_2(x)$ denotes a second-order Legendre polynomial, and 
$P_2^{\prime} = dP_2/dx$ with $x = \mathbf{t}\cdot \mathbf{\hat{r}} = \cos \theta$.
The flow field in the body-fixed frame (bf) can be obtained from that in the lab frame (lf) as 
${\bf u}_\mathrm{bf}({\bf r}) = {\bf u}_\mathrm{lf}({\bf r}) - \mathbf{V} - \bm{\Omega} \times {\bf r}$. 
Note that in Eq.~(\ref{eq:VelocityField}), we have only written terms up to $l = 2$ and do not consider higher-order contributions since they decay more rapidly with $r$. 
Besides, to have a minimal model, we have ignored $l = 2$ modes with $m\neq 0$. 
However, it is straightforward to include the additional terms in the analysis. 
The sign of $\beta_{2 0}^r$ decides the nature of the swimming. 
For example, $\beta_{2 0}^r > 0 $ corresponds to puller-type swimmer and $\beta_{2 0}^r < 0 $ corresponds to pusher-type swimmer (see Fig.~\ref{fig:CS}(b)).
While pullers have an extensile force dipole, resulting, e.g., from the front part of the body, pushers have a contractile force dipole stemming, e.g., from the rear part of the body \cite{Eric_book}.

\section{Hydrodynamic interaction in the presence of lubrication forces}
\label{sec:lub} 

At low Reynolds numbers, microswimmers strongly influence their fluidic environment, as do their nearby swimmers. When the swimmers approach each other, the lubrication forces and torques arise, which play a vital role in the hydrodynamic behavior of the swimmers \cite{wang,yoshinaga,burada_lub}. It has been reported recently that a pair of chiral squirmers portray various behaviors, e.g., divergence, convergence, and even a bounded state \cite{burada} as a result of their mutual hydrodynamic interaction. However, considering the lubrication forces and torques (as in the dense suspension of swimmers), the swimmers do not exhibit convergence behavior \cite{burada_lub}.
In this work, we include the lubrication effects. The lubrication force acting on a swimmer can be calculated by knowing the velocity field 
of the nearby swimmer \cite{wang, yoshinaga}. Due to this lubrication force, the corresponding velocity contribution to a chiral swimmer from the other swimmer, and vice versa, is \cite{burada_lub}
\begin{align}
\label{eq:lub_vel}
U = -2 a^2 B \epsilon \ln \epsilon \,,
\end{align}
where 
$\epsilon$ is half of the separation distance between the swimmers, 
$B = \beta_{10}^{r(1)} t_{13} - \beta_{10}^{r(2)} t_{23}$, 
$\beta_{10}^{r(1)}$ and $\beta_{10}^{r(2)}$ are $l = 1$ and $m = 0$ modes (see the text after Eq.~\ref{eq:slip}) of swimmer one and two, respectively,  
$t_{13}= \mathbf{t}_1 \cdot \mathbf{e}_{_Z}$, 
$t_{23} = \mathbf{t}_2 \cdot \mathbf{e}_{_Z}$, 
$\mathbf{e}_{_Z}$ is the unit vector along the $Z$ direction, and 
$\mathbf{t}_1$ and $\mathbf{t}_2$ are the corresponding orientations of the swimmers 
(see Ref. \cite{burada_lub} for more details). 
Note that the lubrication torques are of the order $O(\epsilon^{1/2})$, 
and which can be neglected in the limit $\epsilon \to 0$. 

The equations of motion of a swimmer in the presence of another swimmer, by considering lubrication forces into account, read as
\begin{align}
\label{eq:gg}
{\bf \dot{q}}_i
  & = \mathbf{V}_i + \mathbf{U}^{\mathrm{lub}}_{i} +\sum^2\limits_{\substack{j=1 \,;\, i\neq j}} \mathbf{u}_j(\mathbf{q}_{ij},\mathbf{n}_{j},\mathbf{b}_{j},\mathbf{t}_{j}) \, \nonumber \\
  \left[\begin{array}{c}  {\bf \dot{n}}_i  \\ {\bf \dot{b}}_i \\{\bf \dot{t}}_i \end{array}\right]  
  & = \left[\bm{\Omega}_i+ \sum^2\limits_{\substack{j=1 \,;\, i\neq j}}\bm{\omega}_j(\mathbf{q}_{ij},\mathbf{n}_{j},\mathbf{b}_{j},\mathbf{t}_{j}) \right] 
  \times   
  \left[\begin{array}{c} {\bf n}_i  \\ {\bf b}_i \\ {\bf t}_i \end{array}\right] \,,
\end{align}
where $\mathbf{U}^\mathrm{lub}_i = U(\cos \theta^\prime \mathbf{t}_i - \sin \theta^\prime \mathbf{n}_i)$ is the additional velocity contribution arising due to the  other swimmer in the lubrication region, 
$\theta^\prime = \cos^{-1}(\mathbf{t}_i \cdot \mathbf{e}_{_Z})$, and the 
vorticity field $\boldsymbol{\omega} = ({\boldsymbol \nabla} \times {\bf u}) /2$. 
Note that $\mathbf{U}^{\mathrm{lub}}_i = 0$ for $R > 2(a + \epsilon)$, and 
$\mathbf{U}_i + \sum^2\limits_{\substack{j=1 \,;\, i\neq j}} \mathbf{u}_j = \sum^2\limits_{\substack{j=1 \,;\, i\neq j}}\bm{\omega}_j= 0$ for 
$R \le 2(a+ \epsilon)$, where 
$\epsilon \ll a$ and $R = |\textbf{q}_{ij}|$ is the radial distance between the swimmers. 

\section{Influence of external linear chemical gradient}
\label{sec:no-noise}

\begin{figure}[t!]
\includegraphics[scale=0.3]{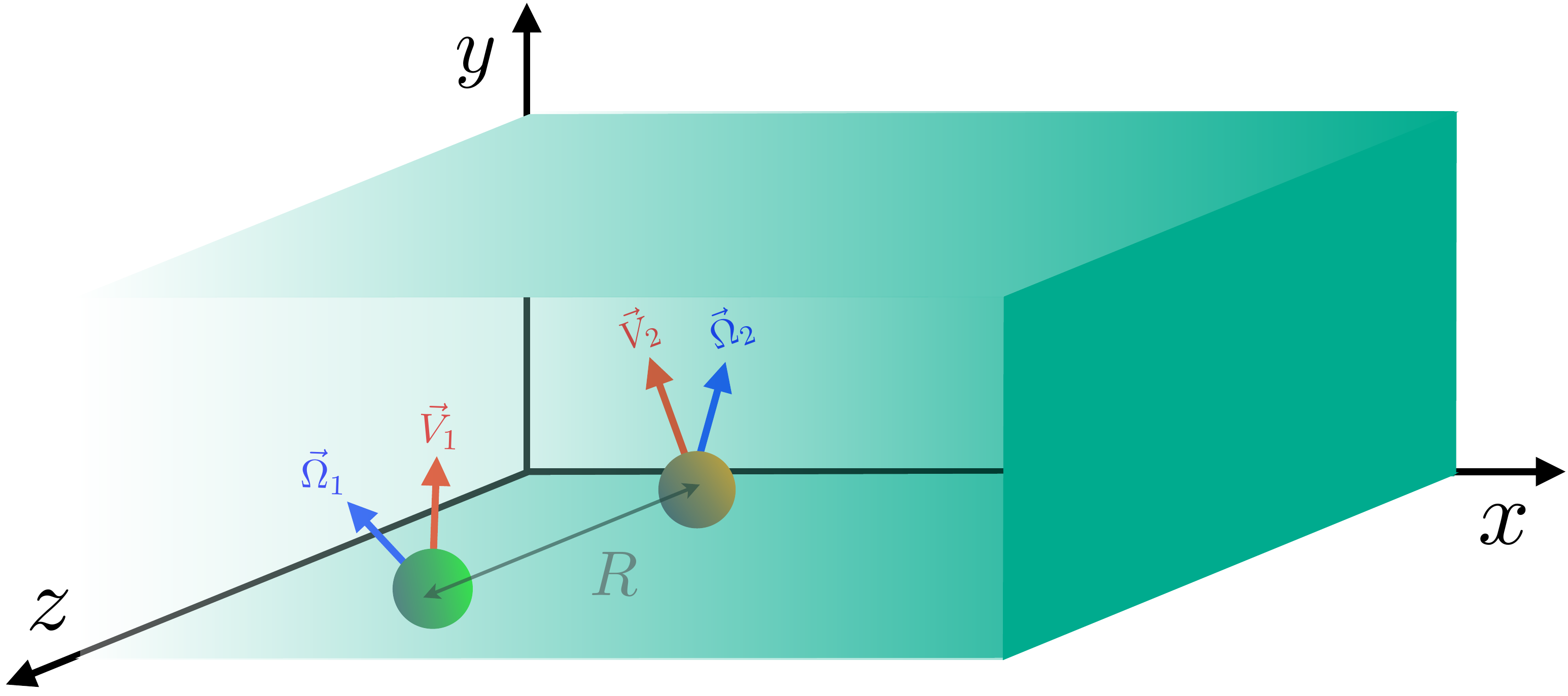}
\caption{\label{fig:sketch}
Schematic diagram of two hydrodynamically interacting chiral swimmers in the presence of a linear chemical gradient applied only along the $x$ direction. The separation distance between the swimmers is $R$.}
\end{figure}

Consider that the two swimmers are subjected to an external chemical concentration field $c(\mathbf{q})$; see the schematic diagram Fig.~\ref{fig:sketch}. As the swimmers sense the gradient, they move from a lower- to a higher-concentration region. Consequently, the local concentration field that the swimmers experience changes with time. The stimulus level ($s = c(\mathbf{q}(t))$) on the swimmer's body is thus a function of time. A swimmer's response to the stimulus is related to the relative change in the chemoattractant concentration ($\Delta c/c$) in the vicinity \cite{ned}. The binding of the chemoattractant to the body's receptors, known as activation, and the unbinding of that, known as deactivation, triggers the internal chemotactic signaling system of the body. 
In our model systems, it results in the modification of the slip coefficients (Eq.~\ref{eq:slip}) of the swimmer, which leads change in the velocity and rotation rate of the swimmer \cite{friedrich, Martin,ruma,burada_jfm}. 

To study the chemotaxis of two swimmers, we use the Barkai–Leibler model \cite{barkai,friedrich,friedrich2} for the adaptation and relaxation mechanism of the swimmer to the external chemical gradient. 
Note that this model is for information processing and memory via the internal biochemical
network, and it has been used widely for bacterial and sperm chemotaxis \cite{Martin,friedrich}. 
The ability of the system to adapt and respond to the external stimulus 
can be captured by a simple dynamical system,
\begin{subequations}
\begin{align}
\begin{split}
\sigma \dot{a_b} &= p_b (s_b + s) - a_b\, ,
\end{split}\\
\begin{split}
\mu \dot{p_b} &= p_b (1-a_b)\, ,
\end{split}
\label{eq:adt_relax}
\end{align}
\label{eq:adt_relax2}
\end{subequations}
where $\sigma$ is the relaxation time, $\mu$ is the adaptation time, 
$a_b(t)$ is the dimensionless output variable, 
$p_b (t)$ is the dynamic sensitivity related to adaptation, and 
$s_b(t)$ arises from the background activity of receptors in the absence of the chemical stimulus. 
The chemoattractant has a dimension of concentration. 
Note that these equations are valid for weak concentration gradients only.
Under a constant stimulus $s(t) = S_c$, 
the system reaches a steady state for which $a_b = 1$ and $p_b = 1/(s_b + S_c)$. 
Therefore, the dynamic sensitivity ($p_b$) maintains an inverse relation with $s_b$ and $s$. 
With increasing stimulus level, $p_b$ decreases. 
The system here is totally adaptive as $a_b(t)$ is independent of $S_c$.

In the presence of an external stimulus, the slip coefficients (Eq.~\ref{eq:slip}) 
of the swimmer are modified as,
\begin{align}
X = X^{(0)} + X^{(1)} \left( a_b(t) - 1 \right) \,, \label{eq:pert:chem}
\end{align}
where
$X =\left \lbrace{\beta_{lm}, \gamma_{lm}}\right \rbrace$,
the unperturbed slip coefficients are 
$X^{(0)} = \left \lbrace{\beta_{lm}^{(0)}, \gamma_{lm}^{(0)}}\right \rbrace$, and the perturbation due to the external gradient is 
$X^{(1)} = \left \lbrace{\beta_{lm}^{(1)}, \gamma_{lm}^{(1)}}\right \rbrace$.
Here, we consider a linear chemical gradient given by \cite{friedrich},
\begin{equation}
c(\mathbf{q}) = c_0 + \mathbf{c_1}\cdot \mathbf{q}\,,
\label{eq:l_gradient}
\end{equation}
where $\mathbf{c_1} = \nabla c(\mathbf{q})= c_1 \mathbf{i}$ is the chemical gradient applied in the $x$ direction and  
$\mathbf{q} = \mathbf{i} x + \mathbf{j} y + \mathbf{k} z$ is the position vector. The chemotactic stimulus is 
$S(t)= c(\mathbf{q(t)})$ \cite{friedrich}.

Using Eq.~(\ref{eq:gg}), we numerically calculate the trajectories of a pair of chiral swimmers, which are hydrodynamically interacting, in the presence of a linear chemical gradient. Note that the velocity and the rotation rate of the chiral squirmer are functions of the slip coefficients, which are affected by the chemical gradient; see Eq. \ref{eq:pert:chem}. As a result, the hydrodynamic interaction between the swimmers is affected. In the present work, we consider chiral swimmers having velocities of equal magnitudes, i.e., $|\mathbf{V}_1| = |\mathbf{V}_2| = v$. However, the rotation rates of the swimmers are in general different and read, $\bm{\Omega_1} = v(\sin\chi_1, 0,\cos\chi_1)/a$ for swimmer one and $\bm{\Omega_2} = v(\sin\chi_2, 0,\cos\chi_2)/a$ for swimmer two. Chemical and hydrodynamic perturbations in the velocity and rotation rate of the swimmers change the corresponding torsion and curvature of the swimmers' helical trajectories. Hence their movements are altered.

As discussed, a swimmer's flow field influences the neighboring swimmer's motion. The hydrodynamic flow field of a swimmer (Eq.~(\ref{eq:VelocityField})) comprises $l = 1, 2$ modes as leading order terms. The $l= 2$ mode is important as it plays a vital role in determining the hydrodynamic interaction between the swimmers. The slip coefficients corresponding to $l = 2$ mode are assumed to have the following form: $3\beta_{20}^{\,r} = 3\gamma_{20}^{\,r} = \lambda_1$ for squirmer one, and $3\beta_{20}^{\,r} = 3\gamma_{20}^{\,r} = \lambda_2$ for squirmer two. The former assumptions are made to minimize the parameters of the problem. Note that for $\lambda_1 \neq \lambda_2$, the flow fields of the swimmers are different. Thus, variation in $\chi_i$ (angle between $\mathbf{V}_i$ and $\bm{\Omega}_i$) and $\pm \lambda_i \,(i = 1,2)$ (the strength of the flow field), quantitatively measure the nature of the interaction between the chiral squirmers and gives rise to several interesting swimming characteristics. Notably, different initial configurations also influence the interaction between the swimmers. We found that, out of various possible initial configurations for the swimmers, swimmers interact with each other for an extended period only in the planar configuration \cite{burada} as is the case in the present study. Here, at time $t = 0$, both the swimmers are on the $xy$ plane, with an initial orientation along the $z-$direction and an initial separation distance $R_0$. 
Our previous study (see Ref.~\cite{burada}) explored the interaction between the two chiral swimmers with the former configuration, which exhibited some new behavior (bounded state) in addition to the ones (convergence and divergence state) displayed by axisymmetric swimmers. We mainly focus on how the external chemical gradient influences these observed behaviors in the present work.

\subsection{State diagram in the $\chi$ - $\lambda$ plane}
\label{subsec:chi-lambda}

\begin{figure*}
\includegraphics[scale=1.175]{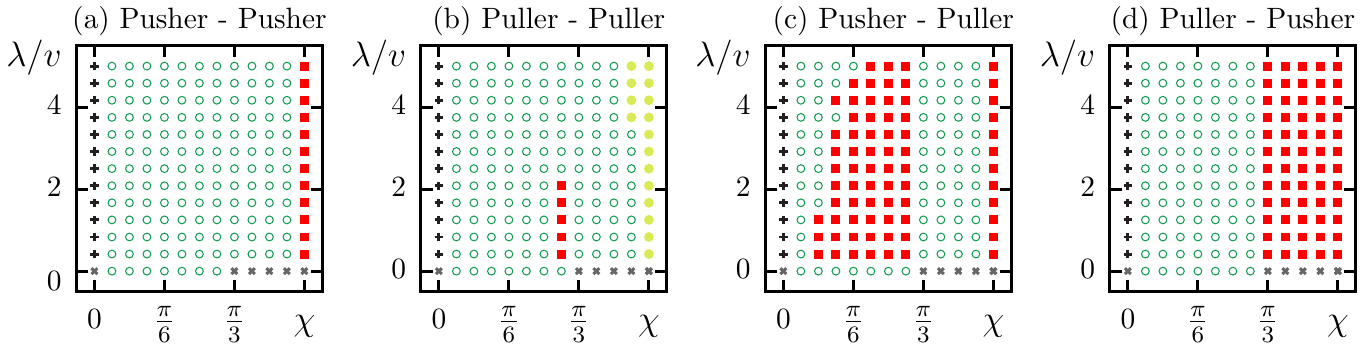}
\caption{(a)-(d) Depicting swimming states of hydrodynamically interacting chiral squirmers performing chemotaxis, where the strength of the chemical gradient $c_1 = 0.1$. 
The symbols imply the following states. Square: bounded state (BS), closed circles:
monotonic divergence state (MD), open circles: divergence state (D), 
plus: locked state, and cross symbols: parallel motion. 
The two swimmers initially start at $\mathbf{q}_1 = (9, 9, 0)a$ and $\mathbf{q}_2 = (3, 3, 0)a$. 
The corresponding initial velocities and rotation rates of the swimmers are 
$\mathbf{V}_1 = \mathbf{V}_2 = v(0, 0, 1)$ and ${\bm \Omega}_1 =
{\bm \Omega}_2 = v (\cos \chi , 0, \sin \chi )/a$, respectively.}
\label{fig:lamchi} 
\end{figure*}

\begin{figure}[htb!]
\includegraphics[scale=0.75]{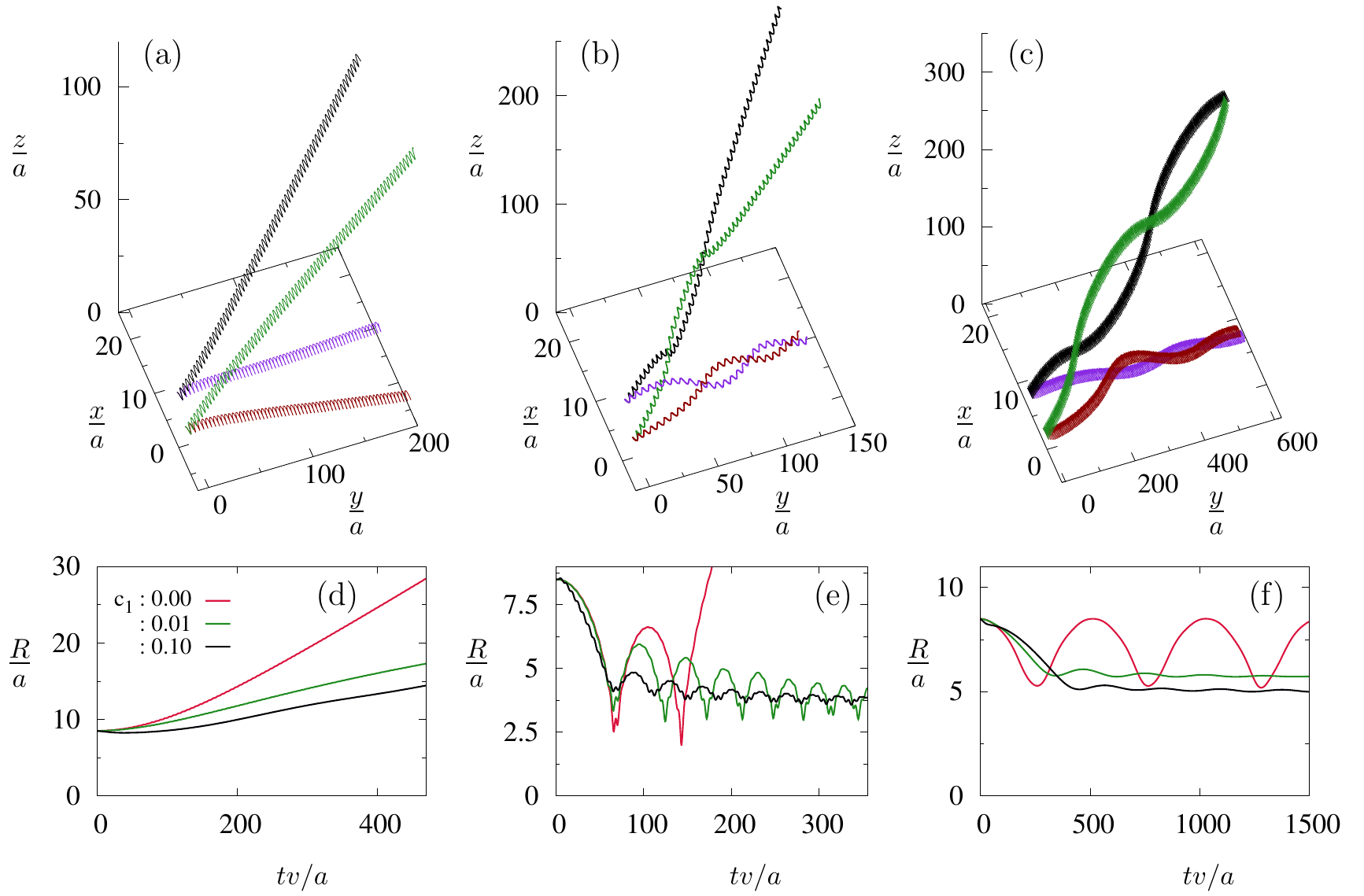}
\caption{\label{fig:traj_lin}
(a)-(c) Show the trajectories of two swimmers in the presence and in the absence of a linear chemical gradient for monotonic divergence (MD), divergence (D), and bounded states (BS).
Here, the strength of the linear gradient $c_1 = 0.05$.
The other parameters are,  
$\chi =  \pi/6, \lambda_1 = -2$, and $\lambda_2 = 2$ for D state,
$\chi =  \pi/3, \lambda_1 = -3, \lambda_2 = 3$ for MD state, and 
$\chi = \pi/3, \lambda_1 = 1, \lambda_2 = -1$ for BS state.
(d)-(f) The corresponding radial separation distance $R$ between the swimmers 
as a function of time. 
Here, lengths are scaled by the radius of the swimmer $a$ and time scaled by $\tau = v/a$.}
\end{figure}

Fig.~\ref{fig:lamchi} depicts the hydrodynamic behavior of two chiral swimmers in the presence of an external chemical gradient. When $\lambda_1 = \lambda_2 = \lambda$ and $\chi_1 = \chi_2 = \chi$, the swimmers are identical (see the Fig.~\ref{fig:lamchi} caption). 
The swimmers portray various behaviors for varying $\lambda/v$ and $\chi$. As reported earlier \cite{burada_lub}, in the absence of a chemical gradient, two chiral swimmers exhibit divergence, monotonic divergence, and bounded states (see Fig.~\ref{fig:traj_lin}) only due to the lubrication effects. These observed states arise due to pure hydrodynamic interaction between the swimmers. However, the chemical gradient converts some monotonic divergence states into divergence states and some divergence states into bounded states (see Fig.~\ref{fig:traj_lin}). Indeed, the chemical stimulus favors the attraction between the swimmers. 
The density of bounded states is more for the odd combination of swimmers like pusher-puller or vice-versa (see fig.~\ref{fig:lamchi}(c) and (d)). 
For other combinations, swimmers rarely exhibit bounded states (see fig.~\ref{fig:lamchi}(a) and (b)). 
The occurrence of bounded motion in even combinations of swimmers is due to the presence of the chemical gradient. Note that in the case of axisymmetric swimmers (without chirality), the hydrodynamic behavior of pusher-puller and puller-pusher-type swimmers is the same. 
However, in the case of chiral swimmers, the hydrodynamic behavior of different swimmers exhibit different behaviors \cite{burada}. It is because the flow pattern of a puller-type chiral swimmer is different from a pusher-type chiral swimmer \cite{burada}. 

Note that for $\chi = 0$ and $\lambda \neq 0$, swimmers move in a straight line in the $z$ direction, attract each other, and converge to a locked state \cite{burada_lub}. In this state, the swimmers approach a minimum distance and stay together. For $\chi = 0$ (straight line motion) and $\chi = \pi/2$ (circular motion), the neutral swimmers, i.e., $\lambda = 0$, move parallel to each other without influencing each other \cite{burada}. Not only the chemical gradient but the initial separation distance $R_0$ (at $t = 0$) between the swimmers also plays a vital role in the hydrodynamic interaction between the swimmers. Interestingly, as $R_0$ increases, the bounded state is more favorable. It is reported in detail in our earlier studies (see Refs.~\cite{burada,burada_lub}). 

We also study the interaction between non-identical swimmers, i.e., $\lambda_1 \neq \lambda_2$ and $\chi_1 \neq \chi_2$ (see Appendix.\ref{subsec:lam-lam}). For a fixed $\chi = \pi/3$ and varying flowfield strengths $(\lambda_1, \lambda_2)$, we observe mainly divergence states and a few bounded states for the different combinations of the swimmers (see fig.~\ref{fig:lamlam}(a)). 
Identical or nearly identical swimmers mostly exhibit bounded motion primarily for the puller-pusher combination 
(see fig.~\ref{fig:lamlam}(a), top right panel). 
However, other combinations also exhibit bounded motion when $\lambda_1 = \lambda_2 = 0$ (see fig.~\ref{fig:lamlam}(a)).
The former implies that non-identical field strength is unfavorable for the bounded motion. 
On the other hand, we can fix the flowfield strengths at $\lambda = v$ and vary relative orientations of the swimmers $(\chi_1, \chi_2)$ (see fig.~\ref{fig:lamlam}(b)). Non-identical orientations are associated with divergence and monotonic divergence states. See Appendix \ref{subsec:lam-lam} for a detailed discussion.

Fig.~\ref{fig:traj_lin} depicts the trajectories and the radial separation distance between the swimmers in the presence and absence of the chemical gradient. As mentioned before, the simultaneous effect of the hydrodynamic interaction and the chemical stimulus converts some of the divergence states into bounded states (see figs. \ref{fig:traj_lin} (b),(e)). 
Here, the stimulus increases the attraction between the swimmers. 
As a result, for the bounded state, the swimmers tend to align in the direction of the chemical gradient. The oscillation in the separation distance ($R$) between the swimmers decreases (see figs.~\ref{fig:traj_lin}(b),(c),(e),(f)) when compared to the stimulus-free situation. Whereas for the monotonic divergence state, the swimmers diverge at a slower rate than the stimulus-free case (see figs. \ref{fig:traj_lin} (a),(d)). As reported earlier, in the case of a single chiral swimmer, the body completely aligns its helix axis to the direction of the gradient \cite{ruma}. In the two-swimmer case, none of the swimmers can make their helix axis parallel to the direction of the gradient because of the hydrodynamic flow field generated by the neighboring swimmer. In particular, it is due to the dominating term in the hydrodynamic flow field $o(1/r^2)$ and the weak chemical gradient we apply.

\section{Dependence of relative success time on chemical gradient}

\label{sec:successtime}
\begin{figure}
\includegraphics[scale = 0.8]{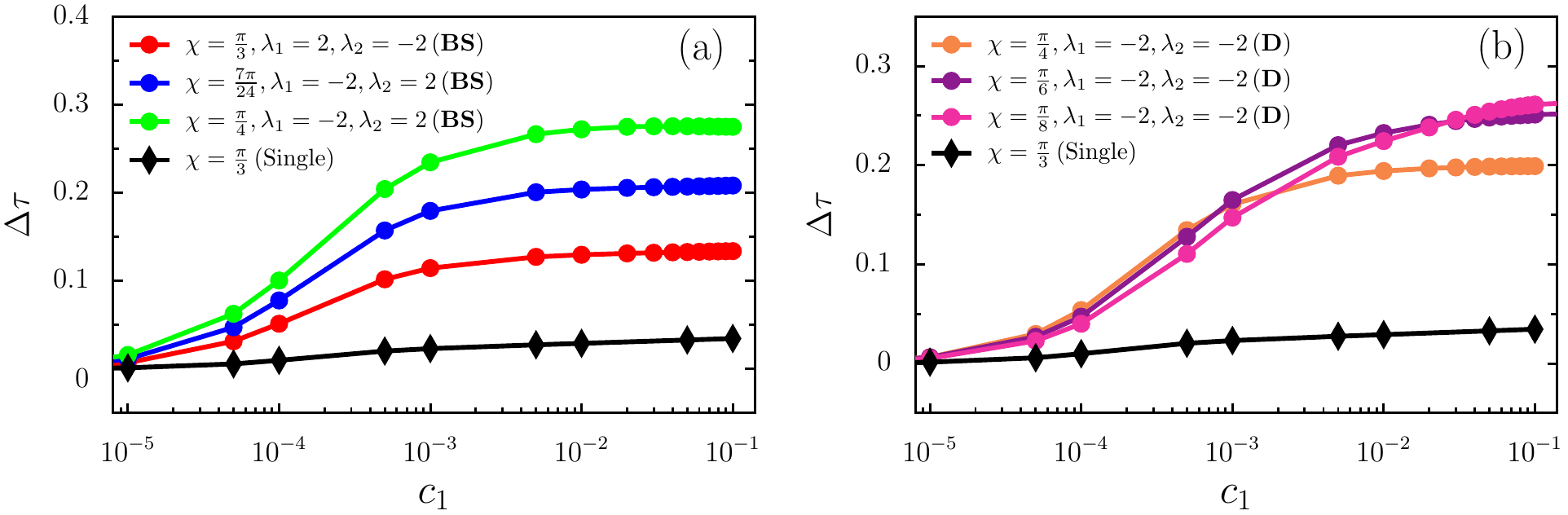}
\caption{\label{fig:success}  
The relative success time $\Delta \tau = (t^0 - t^c)/t^0$ as a function of the chemical gradient strength $c_1$ for various observed states, (a) for bounded states and (b) for divergence states. The diamond symbols for the single swimmer. Here, lines connect the data points.
}
\end{figure}

A question may arise whether a pair of swimmers or a single swimmer is efficient in swimming in the direction of the external chemical gradient. To answer this, we measure the relative success time defined as $\Delta \tau = (t^0 - t^c)/t^0$. Here, $t^0$ is the time a pair of swimmers or a single swimmer takes to reach a particular reference point on the $x$ axis in the absence of the chemical gradient, and $t^c$ is the same in the presence of the chemical gradient. Note that $\Delta \tau$ increases with $c_1$ and saturates for higher $c_1$. The latter is due to the saturation of the internal signaling network of the bodies \cite{friedrich2}. 
Henceforth, the network cannot modify its translational and rotational velocities further.

We find pair swimming is more efficient than individual swimming (see Fig. \ref{fig:success}). It is because the combined effect of the hydrodynamic interaction between the swimmers and chemical perturbation increases the motility of the swimmers. Interestingly, the bounded swimmers are more efficient than others (see Fig. \ref{fig:success}(a)), especially for lower values of $\chi$. However, the influence of $\chi$ on chemotactic success rate is minimal for diverging motion (see Fig. \ref{fig:success}(b)). Note that, the range of $c_1$ is from $o(10^{-6})$ to $o(10^{-1})$. For weak gradients, divergence motion persists. With increasing strength of $c_1$, as mentioned earlier, some divergence states get converted into bounded states. Note that for values of $\chi \sim \pi/2$, swimmers effectively move in circular paths, and correspondingly the success rate is low (not shown). 

\section{Conclusions}
\label{sec:conclusions}

In this work, we have studied the combined behavior of two chiral swimmers by considering the lubrication effect in the presence of a linear chemical gradient. We have used a generalized squirmer model called the chiral squirmer. In general, a pair of chiral swimmers can exhibit bounded, divergence, convergence, and monotonic convergence states without a chemical gradient. However, only bounded, divergence, and monotonic divergence states are observed in the current study. The convergence and monotonic convergence states are absent because of the lubrication force between the swimmers. Due to this lubrication force, swimmers repel each other when they approach each other.

Interestingly, a chemical gradient converts some divergence- and monotonic-divergence states into bounded states. The chemical gradient favors the attraction between the swimmers. It leads to efficient swimming. We have presented detailed state diagrams depicting the hydrodynamic behavior of two chiral swimmers in the presence of a chemical gradient. We have computed the relative success time of a pair of swimmers in aligning their movement toward the chemical gradient, whose strength is controlled by the parameter $c_1$. We find that pair swimming is more efficient than individual swimming as the combined effect of the hydrodynamic interaction between the two swimmers and chemical perturbation increases the motility of the swimmers. Interestingly, the bounded swimmers are more efficient than others.

This study is not limited to chemotaxis but can be generalized to various external gradients, e.g., phototaxis and gyrotaxis. For these cases, one needs to identify how the slip coefficients of the chiral swimmer can be modified according to the external gradients. Also, the present study helps design artificial micro locomotors which can be used for drug delivery and targeted applications.

\section{Acknowledgements}
The authors acknowledge the financial support of the Indian Institute of Technology Kharagpur, India.

\section{Data Availability}
The data supporting this study's findings are available from the corresponding author upon reasonable request.

\appendix
\section{State diagrams of $\lambda_1$ - $\lambda_2$ and $\chi_1$ - $\chi_2$}
\label{subsec:lam-lam}

\begin{figure*}
\includegraphics[scale=1]{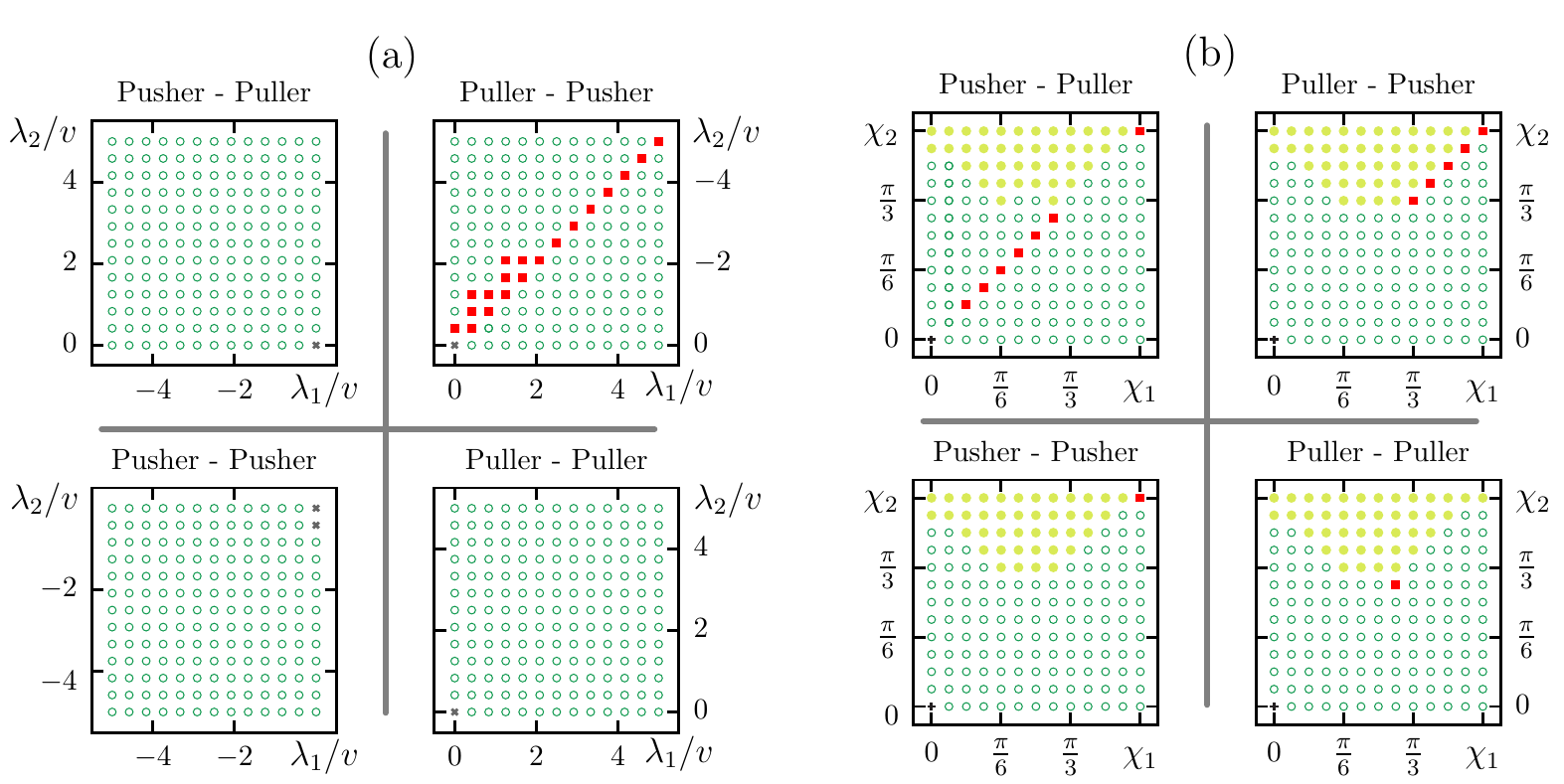}
\caption{\label{fig:lamlam} 
State diagrams corresponding to two hydrodynamically interacting chiral swimmers in the presence of chemotaxis. (a) For varying field-strengths $(\lambda_1, \lambda_2)$ and (b) for $(\chi_{_1}, \chi_{_2})$. See the main text for details. The initial conditions and symbol descriptions are the same as in figure~\ref{fig:lamchi}. We set $\chi_1 = \chi_2 = \pi/3$ for the left panel and $|\lambda_1| = |\lambda_2| = |\mathbf{V}| = v$ for the right panel.}
\end{figure*}

Fig.~\ref{fig:lamlam}(a) depicts the $\lambda_1-\lambda_2$ state diagram of two chiral swimmers, in the presence of a chemical gradient, for fixed $\chi_1$ and $\chi_2$. The additive contributions of hydrodynamic interaction and chemical perturbations drive the swimmers away from each other. Only the swimmers with identical or nearly identical field strengths $(|\lambda_1| = |\lambda_2|)$ exhibit bounded motion (see Fig.~\ref{fig:lamlam}(a)). Moreover, identical combinations of swimmers, i.e., puller-puller or pusher-pusher, mostly display a divergence state 
with a rare occurrence of bounded states. The probability of observing a bounded state is more if the first swimmer is a puller and the second swimmer is a pusher. However, the opposite combination does not work while $|\lambda_1| \neq 0, |\lambda_2| \neq 0$. The reason is the asymmetry in the swimmer's flow field about the orientation of the swimmer. At different positions around a chiral swimmer, the neighboring swimmer experiences different hydrodynamic forces. As a result, the interaction between the swimmers is different for their different relative positions (see Ref.~\cite{burada} for more details).

Fig.~\ref{fig:lamlam}(b) shows the $\chi_1-\chi_2$ state diagram of two chiral swimmers, in the presence of a chemical gradient, for fixed $\lambda_1$ and $\lambda_2$. The swimmers do not exhibit bounded motion unless $\chi_1 = \chi_2$. As mentioned before, similar combinations of the swimmers, i.e., puller-puller or pusher-pusher, exhibit bounded states rarely. However, the opposite combinations, i.e., puller-pusher or pusher-puller, do exhibit more bounded states (see Fig.~\ref{fig:lamlam} (b)). In the absence of a chemical gradient, more monotonic divergence states are observed (see Ref.~\cite{burada}). However, the presence of the chemical field increases the attraction between the swimmers and converts some of the monotonic divergence (MD) states into divergence (D) states (see Fig.~\ref{fig:lamlam}(b)).

\bibliography{chemotaxis_pof_v1}

\end{document}